\def\CC{{\mathchoice
{\rm C\mkern-8mu\vrule height1.45ex depth-.05ex 
width.05em\mkern9mu\kern-.05em}
{\rm C\mkern-8mu\vrule height1.45ex depth-.05ex 
width.05em\mkern9mu\kern-.05em}
{\rm C\mkern-8mu\vrule height1ex depth-.07ex 
width.035em\mkern9mu\kern-.035em}
{\rm C\mkern-8mu\vrule height.65ex depth-.1ex 
width.025em\mkern8mu\kern-.025em}}}

\def\RR{{\rm I\kern-1.6pt {\rm R}}}

\def\ZZ{{\rm Z}\kern-3.8pt {\rm Z} \kern2pt}

\input phyzzx.tex

\def\np{Nucl. Phys.}
\def\pl{Phys. Lett.}

\def\ap{Ann. Phys.}
\def\cmp{Comm. Math. Phys.}
\def\jmp{J. Math. Phys.}
\def\ijmp{Int. J. Mod. Phys.}

\def\lmp{Lett. Math. Phys.}

\def\faa{Funct. Anal. and Appl.}

\tolerance=500000
\overfullrule=0pt
\Pubnum={US-FT-31/96\cr hep-th/9606180}
\pubnum={US-FT-31/96}
\date={June, 1996}
\pubtype={}
\titlepage

\title{On the free field realization of the ${\rm osp}(1\vert 2)$ 
 current algebra} 
\author{I.P. Ennes\foot{E-mail: ENNES@GAES.USC.ES}, 
A.V. Ramallo\foot{E-mail: ALFONSO@GAES.USC.ES} and 
J. M. Sanchez de Santos \foot{E-mail: SANTOS@GAES.USC.ES} }
\address{Departamento de F\'\i sica de
Part\'\i culas, \break Universidad de Santiago, \break
E-15706 Santiago de Compostela, Spain. \break}

\abstract{The free field representation of the
 ${\rm osp}(1\vert 2)$ current algebra is analyzed. The four point
conformal blocks of the theory are studied. The structure
constants for the product of an arbitrary primary operator and a
primary field that transforms according to the fundamental
representation of ${\rm osp}(1\vert 2)$ are explicitly
calculated.}

\endpage
\pagenumber=1
\sequentialequations

\hyphenation {o-pe-ra-tor}
The free field constructions are a powerful tool in conformal
field theory (CFT). The Feigin-Fuchs formalism
\REF\FF{B. L. Feigin and D. B. Fuchs,
\journal\faa&13, No.4(79)91\journal\faa&16, No.2(82)47.}[\FF] has
allowed to study the conformal block structure of the minimal
models and to obtain the operator algebra of their primary fields
\REF\DF{Vl.S.Dotsenko and V. A. Fateev \journal\np&B240(84)312
\journal\np&B251(85)691\journal\pl&B154(85)291.} [\DF]. 
The extension of the Feigin-Fuchs approach to the minimal
superconformal models has been performed in refs.
\REF\SCFT{H. Eichenherr \journal\pl&B151(85)26; M.A. Bershadsky, 
V. G. Knizhnik and M. G. Teitelman \journal\pl&B151(85)31;
D. Friedan, Z. Qiu and S. Shenker \journal\pl&B151(85)37.}
\REF\Kita{Y. Kitazawa et al. \journal\np&B306(88)425.}
\REF\Zaugg{L. Alvarez-Gaume and P. Zaugg \journal\ap&215(92)171.}
[\SCFT, \Kita, \Zaugg]. In the case
of current algebras the Wakimoto representation 
\REF\waki{M. Wakimoto\journal\cmp&104(86)604.}
\REF\gera{A. Gerasimov et al. \journal\ijmp&A5(90)2495.}
\REF\Feigin{B. L. Feigin and E. V. Frenkel in ``Physics and
Mathematics of Strings", edited by L. Brink et al., World
Scientific, 1990 \journal\cmp&128(90)161 
\journal\lmp&19(90)307.} 
[\waki, \gera, \Feigin] plays a similar
r\^ole. Indeed in ref. 
\REF\Dotsenko{Vl. S.
Dotsenko\journal\np&B338(90)747\journal\np&B358(91)547.} 
[\Dotsenko] Dotsenko has presented an analysis of the
free field representation of the $SU(2)$ CFT which leads to a direct
evaluation of the correlation functions and structure constants of
the model.

It is the purpose of this paper to perform a similar study for the
case of a CFT enjoying an affine symmetry based on the Lie
superalgebra ${\rm osp}(1\vert 2)$
\REF\scheunert{For an account of the general theory of Lie
superalgebras see M. Scheunert,  ``The Theory of Lie
Superalgebras",  {\sl Lect. Notes in Math.} 716,
Springer-Verlag, Berlin (1979).}
[\scheunert]. To motivate our work let us
mention that the ${\rm osp}(1\vert 2)$ theory is related to the
$N=1$ superconformal minimal models via the quantum hamiltonian
reduction 
\REF\bershadsky{M. Bershadsky and H.
Ooguri\journal\pl&B229(89)374.}
[\bershadsky]. Moreover, the topological version of the 
${\rm osp}(1\vert 2)$ CFT, \ie\ the
${\rm osp}(1\vert 2)/{\rm osp}(1\vert 2)$ coset model, is related
to the non-critical Ramond-Neveu-Schwarz superstring
\REF\yu{J. B. Fan and M. Yu, ``G/G Gauged
Supergroup Valued WZNW Field Theory", Academia Sinica preprint
AS-ITP-93-22, hep-th/9304123.}
\REF\Ennes{I. P. Ennes, J. M. Isidro and A. V.
Ramallo\journal\ijmp&A11(96)2379.}
[\yu, \Ennes]. We
have undertaken this study with the hope that a complete solution
of the ${\rm osp}(1\vert 2)$ theory could make more explicit these
connections.

The ${\rm osp}(1\vert 2)$  current algebra is generated by 
three bosonic currents $J_{\pm}$ and $H$, together with two
fermionic currents $j_{\pm}$. This algebra can be realized
[\bershadsky] in terms of one scalar field $\phi$, two conjugate
bosonic fields $(w, \chi)$ and one fermionic $bc$ system (denoted
by $(\bar \psi, \psi)$). The  dimensions of these fields are
$\Delta(w)=\Delta (\bar\psi)=1$ and
$\Delta(\chi)=\Delta (\psi)=0$. The basic operator product
expansions (OPE's) will be taken as:
$$
w(z_1)\,\chi(z_2)\,=\,\psi(z_1)\,\bar\psi(z_2)\,=\,{1\over z_1-z_2}
\,\,\,\,\,\,\,\,\,\,\,\,\,\,\,\,\,\,
\phi(z_1)\,\phi(z_2)\,=\,-{\rm log}\,(z_1-z_2)\,.
\eqn\uno
$$
In terms of these fields the ${\rm osp}(1\vert 2)$ currents can be
represented as:
$$
\eqalign{
J_+\,=&\,w\cr
J_-\,=&-\,w\chi^2\,+\,i\sqrt{2k+3}\,\,\chi\partial\phi\,-
\,\chi\psi\bar\psi\,+k\partial\chi\,+
\,(k+1)\psi\partial\psi\cr
H\,=&-w\chi\,+{i\over 2}\,\sqrt{2k+3}\,\,\partial\phi\,-\,
{1\over 2}\,\psi\bar\psi\cr
j_+\,=&\bar\psi\,+\,w\psi\cr
j_-\,=&-\chi(\bar\psi\,+\,w\psi)\,+i\sqrt{2k+3}\,\,
\psi\partial\phi\,+\,(2k+1)\partial\psi\,.\cr}
\eqn\dos
$$
In eq. \dos\ the c-number $k$ is the level of the affine 
${\rm osp}(1\vert 2)$ superalgebra. In this paper we shall restrict
ourselves to the case in which $k$ is a positive integer. Recall
that
$k$ parametrizes the central extension of the algebra. As a CFT, 
the 
${\rm osp}(1\vert 2)$ current algebra is characterized by its
energy-momentum tensor $T$. The explicit form of $T$ in terms of the
currents can be obtained from the Sugawara construction. Using 
the representation \dos\ one can easily get  $T$ as a function  
of the free fields:
$$
T\,=\,w\partial\chi\,-\,\bar\psi\partial
\psi\,-\,{1\over 2}\,(\partial\phi)^2\,+\,
{i\over 2}\,\alpha_0\,\partial^2\phi\,.
\eqn\tres
$$
In \tres\ $\alpha_0$ is a background charge, whose expression in
terms of the level $k$ is given by:
$$
\alpha_0\,=\,-{1\over \sqrt{2k+3}}\,.
\eqn\cuatro
$$
It is easy to verify from equations \tres\ and \cuatro\ that $T$
satisfies the Virasoro algebra with central charge
$c\,=\,{2k\over 2k+3}$.

The irreducible representations of the ${\rm osp}(1\vert 2)$
 superalgebra have been studied in ref.
\REF\Pais{A. Pais and V. Rittenberg\journal\jmp&16(75)2063; 
M. Scheunert, W. Nahn and  V. Rittenberg\journal\jmp&18(77)155.}
[\Pais]. They are characterized
by their isospin $j$, which can be either integer or half-integer,
and the statistics of their highest weight states. We shall denote
a general state of the isospin $j$ multiplet by $|j,m>$, where
$m$ is the eigenvalue of the generator $H$  
($m=-j, -j+{1\over 2},\,\cdots\,,j-{1\over 2}, j$). Obviously
$|j,j>$ is the highest weight state and the dimensionality of 
the representation is $4j+1$. If $|j,j>$ is bosonic
(fermionic) we will say that the representation is even (odd).
Notice that when  $j-m$ is integer (half-integer) the states $|j,m>$
and $|j,j>$ have the same (opposite) statistics. In the affine
theory we have a primary field $\Phi^j_m$ associated to every state 
 $|j,m>$. It is not difficult to obtain a representation of 
$\Phi^j_m$ within our free field realization. One can check that
the operators
$$
\Phi^j_m\,=\,\cases{\chi^{j-m}\,e^{-2ij\alpha_0\,\phi}
                     &if $j-m\in \ZZ$\cr\cr
                    \chi^{j-m-{1\over 2}}\,\psi\,
                    e^{-2ij\alpha_0\,\phi}
                    &if $j-m\in \ZZ\,+{1\over 2}\,$,}
\eqn\cinco
$$
satisfy all the requirements. First of all it is easy to check that
they are primary operators with a conformal weight which is
independent of $m$ and given by:
$$
\Delta_j\,=\,{j(2j+1)\over 2k+3}\,.
\eqn\seis
$$
Moreover, the quantity $m$ is precisely the eigenvalue of the
generator $H$, as it is evident from the OPE
$$
H(z_1)\,\Phi^j_m(z_2)\,=\,m\,\,{\Phi^j_{m}(z_2)\over z_1-z_2}\,.
\eqn\siete
$$
Acting with the $J_{\pm}$ operators we change the value of $m$ by
one unit
$$
J_{\pm}(z_1)\,\Phi^j_m(z_2)\,=\,\cases{
             (j\mp m)\,\,{\Phi^j_{m\pm 1}(z_2)\over z_1-z_2}
             &if $j-m\in \ZZ$\cr\cr\cr
             (j\mp m\,-{1\over 2})\,\,{\Phi^j_{m\pm 1}(z_2)
              \over z_1-z_2}
             &if $j-m\in \ZZ\,+{1\over 2}\,$,}
\eqn\ocho
$$
whereas the fermionic currents $j_{\pm}$ generate a shift of $m$ by
one-half unit
$$
j_{\pm}(z_1)\,\Phi^j_m(z_2)\,=\,\cases{
             (j\mp m)\,\,{\Phi^j_{m\pm 1/2}(z_2)\over z_1-z_2}
             &if $j-m\in \ZZ$\cr\cr\cr
             \pm\,\,{\Phi^j_{m\pm 1/2}(z_2)\over z_1-z_2}
             &if $j-m\in \ZZ\,+{1\over 2}\,$.}
\eqn\nueve
$$
The screening operators are a basic ingredient in any 
free field representation. These are operators that (anti)commute
with all the currents. They can be obtained by integrating a local
operator over a closed contour:
$$
Q\,=\,\oint\,dz\,S(z)\,,
\eqn\diez
$$
where $S(z)$ has conformal weight equal to one and is
such that its OPE's with the ${\rm osp}(1\vert 2)$ currents have
only total derivatives. This last condition guarantees the
(anti)commutation of $Q$ with $J_{\pm}$, $H$ and $j_{\pm}$. It is
easy to verify[\bershadsky] that $S$ can be taken as:
$$
S\,=\,(\,\bar\psi\,-\,w\psi\,)\,e^{i\alpha_0\phi}\,.
\eqn\once
$$
In order to get integral representations for the conformal blocks
of the theory one needs to construct conjugate representations for
the primary fields. Let us proceed as it was done in ref.
[\Dotsenko] for the $SU(2)$ case. First of all, we ask ourselves
which is the operator conjugate to the identity. This operator
$\tilde I$ must have conformal dimension $\Delta\,=\,0$ and, after
modding out spurious states, its OPE's with the currents must only
contain regular terms. These conditions allow to determine the form
of
$\tilde I$. One gets:
$$
\tilde I\,=\,w^s\,e^{2is\alpha_0\phi}\,,
\eqn\doce
$$
where $s\,=\,-k-1$. The form of $\tilde I$ given in equation
\doce\ fixes the charge asymmetry conditions that one must impose
to any correlator in order to get a non-vanishing result. Let us
suppose that we are computing the vacuum expectation value 
$<\,\prod_{i}\,O_i\,>$ where $O_i$ are general operators of 
the form $O_i\,=\,w^{n_i}\,\chi^{m_i}\,e^{i\alpha_i\phi}\,$. Calling 
$N(w)\,=\,\sum_i\,n_i$ and $N(\chi)\,=\,\sum_i\,m_i$, one gets
after inspecting eq. \doce,  the following conditions:
$$
\eqalign{
&N(w)\,-\,N(\chi)\,=\,s\cr
&\sum_i\,\alpha_i\,=\,2\alpha_0 s\,.\cr}
\eqn\trece
$$
As it happened for the $SU(2)$ case (see [\Dotsenko]) it is simple
to obtain the expression of the operator $\tilde \Phi_j^j$ conjugate
to the highest weight primary field. The result is:
$$
\tilde \Phi_j^j\,=\,w^{2j+s}\,\,
e^{2i(j+s)\alpha_0\,\phi}\,.
\eqn\catorce
$$
Acting on $\tilde \Phi_j^j$ with $j_-$ and $J_-$ one gets the
explicit form of the remaining  operators $\tilde \Phi_j^m$ of
the conjugate multiplet. The expressions  obtained  in this way
become increasingly involved as $m$ is decreased. To illustrate
this fact, let us write the simplest of those fields:
$$
\tilde \Phi_{j-1/2}^j\,=\,{1\over 2j}\,[\,
(2j+s)\,\bar\psi\,w^{2j+s-1}\,-\,s\,w^{2j+s}\,\psi\,]\,
e^{2i(j+s)\,\alpha_0\phi}\,.
\eqn\quince
$$

We shall try now to apply the general formalism developed to the
study of the four point conformal blocks of the model. By using
the $SL(2,\ZZ)$ projective invariance of the Virasoro algebra one
can fix the positions of the four fields involved in the
correlator to the values $z_1\,=\,0$, $z_2\,=\,z$, $z_3\,=\,1$
and $z_4\,=\,\infty$. In general we will have to study a vacuum
expectation value of the
form $<\,\Phi^{j_1}_{m_1}(0)\,\Phi^{j_2}_{m_2}(z)\,
\Phi^{j_3}_{m_3}(1)\,\tilde\Phi^{j_4}_{m_4}(\infty)\,
Q^n\,>$, where the four primary fields of the correlator will be 
taken to correspond to even ${\rm osp}(1\vert 2)$
representations. A simple counting using 
the second selection rule in
\trece\ fixes the number of screening operators that one must
insert in the correlator to the value 
$n\,=\,2\,(\,j_1\,+\,j_2\,+\,j_3\,-\,j_4\,)$.  We shall restrict
ourselves to the case in which $j_3\,=\,j_2$ and 
$j_4\,=\,j_1$ with $j_1\geq j_2$. Notice that in this case 
$n\,=\,4j_2$, which in particular means that the number of
screening operators must be even. Let us restrict further the
values of the $m$'s to be $m_1\,=-j_1$, $m_2\,=\,j_2$, $m_3\,=-j_2$
and $m_4\,=\,j_1$. It can be seen that the study of this particular
case is enough to determine the operator algebra of the model.
Accordingly, let us define the quantity:
$$
I(z)\,\equiv\,
<\,\Phi^{j_1}_{-j_1}(0)\,\Phi^{j_2}_{j_2}(z)\,
\Phi^{j_2}_{-j_2}(1)\,\tilde\Phi^{j_1}_{j_1}(\infty)\,
Q^{4j_2}\,>\,.
\eqn\dseis
$$
Using eqs. \cinco, \catorce, \diez\ and \once\ it is
straightforward to get the following representation for $I(z)$:
$$
I(z)\,=\,
\prod_{i=1}^{n}\,\,\oint_{C_i}\,\,dt_i\,
\lambda(z,\{t_i\})\,\eta(\{t_i\}),
\eqn\dsiete
$$
where $C_i$ are certain integration contours to be specified, the
function $\lambda(z,\{t_i\})$ is given by:
$$
\eqalign{
\lambda(z,\{t_i\})\,=\,
<\,e^{-2ij_1\alpha_0\,\phi(0)}\,e^{-2ij_2\alpha_0\,\phi(z)}\,
e^{-2ij_2\alpha_0\,\phi(1)}\,
e^{2i(s+j_1)\alpha_0\,\phi(\infty)}\,
e^{i\alpha_0\,\phi(t_1)}\cdots e^{i\alpha_0\,\phi(t_n)}\,>\cr},
\eqn\docho
$$
and $\eta(\{t_i\})$ is:
$$
\eqalign{
\eta(\{t_i\})\,=&\,(-1)^{2j_2}\,
<\,(\chi(0))^{2j_1}\,(\chi(1))^{2j_2}\,
(w(\infty))^{2j_1+s}\,w(t_1)\,\cdots\,w(t_{2j_2})\,\,>\times\cr\cr
&\times\,<\,\psi(t_1)\cdots\psi(t_{2j_2})\,
\bar\psi(t_{2j_2+1})\cdots\bar\psi(t_{4j_2})\,>\,+\,
{\rm permutations.}\cr}
\eqn\dnueve
$$
In \dnueve\ we have taken into account the fact that, according to
the first condition in \trece, one needs to pick up $2j_2\,$ $w$
fields from the screening charges. The sum over permutations
indicated in eq. \dnueve\ includes all the possible ways to choose 
$2j_2\,$  $w$ fields from the $4j_2\,$ screening operators. Notice
that a different fermionic correlator is associated to each of
these elections.

Let us now specify the contours appearing in \dsiete. We shall use
the canonical set of integrations that correspond to the s-channel
conformal blocks [\DF, \Dotsenko]. We will take 
the first $n-p+1$ integrals
along a path lying in the real axis and joining the points $t=1$
and $t=\infty$, where $1\leq p\leq n+1 $. 
On the other hand, the remaining $p-1$ integrals
will be taken between the points $t=0$ and $t=z$.  The
integrations in the intervals $(1,\infty)$ and $(0,z)$ will be
considered as ordered integrations. Therefore, denoting 
the holomorphic conformal block corresponding to the 
contours just described
by $I_p(z)$, we can write: 
$$
I_p(z)\,=\,\int_1^{\infty}\,du_1\cdots\int_1^{u_{n-p}}\,
du_{n-p+1}\int_0^{z}\,dv_1\cdots\int_0^{v_{p-2}}\,dv_{p-1}\,
\lambda_p(z,\{u_i\},\{v_i\})\,\eta_p(\{u_i\},\{v_i\}),
\eqn\veinte
$$
where we have relabelled the integration variables $t_i$ as 
$u_i=t_i$ for $i=1,\cdots, n-p+1$ and $v_i=t_{n-p+1+i}$ for 
$i=1,\cdots, p-1$. The quantity $\eta_p(\{u_i\},\{v_i\})$
is the function  $\eta(\{t_i\})$ after this relabelling. Similarly
one can get the expression of $\lambda_p(z,\{u_i\},\{v_i\})$ from
eq. \docho. Indeed, by applying Wick theorem to the vacuum
expectation value of exponentials displayed in eq. \docho, one
gets:
$$
\eqalign{
\lambda_p(z,\{u_i\},\{v_i\})\,=&\,z^{8j_1j_2\rho}\,
(1-z)^{8j_2^2\rho}\,
\prod_{i=1}^{n-p+1}\,u_i^a\,(u_i-z)^b\,(u_i-1)^b\,
\prod_{i<j}(u_i-u_j)^{2\rho}
\times\cr
&\times
\prod_{i=1}^{p-1}\,v_i^a\,(z-v_i)^b\,(1-v_i)^b
\prod_{i<j}(v_i-v_j)^{2\rho}
\,\prod_{i=1}^{n-p+1}\,\prod_{j=1}^{p-1}
(u_i-v_j)^{2\rho},\cr}
\eqn\vuno
$$
where we have defined
$$
\rho\,=\,\alpha_0^2/2\,=\,{1\over 2(2k+3)}\,,
\,\,\,\,\,\,\,\,\,\,\,\,\,\,
a\,=\,-2j_1\alpha_0^2\,,
\,\,\,\,\,\,\,\,\,\,\,\,\,\,
b\,=\,-2j_2\alpha_0^2\,.
\eqn\vdos
$$

The physical correlation function $G(z\,,\,\bar z)$ is obtained by
combining, in a monodromy invariant form, holomorphic and
antiholomorphic blocks:
$$
G(z\,,\,\bar z)\,=\,\sum_p\,X_p\,|\,I_p(z)\,|^2\,,
\eqn\vtres
$$
where the constants $X_p$ will be specified later.

A simple analysis of the representation given in eq. \veinte\
allows to determine the non-analytic behaviour of the functions 
$I_p(z)$ around the point $z=0$. The result is:
$$
I_p(z)\,\sim\,N_p\,z^{\Delta_r\,-\,
\Delta_{j_1}\,-\,\Delta_{j_2}}\,,
\eqn\vcuatro
$$
where $N_p$ is a constant and the exponent $\Delta_r$ is the
conformal weight of  a primary field of isospin 
$r\,=\,j_1\,+\,j_2\,+{1\,-\,p\over 2}$. This number $r$ has the
interpretation of the isospin of the s-channel intermediate
state. In order to confirm this identification, let us notice
that as $p\,=\,1,\cdots,4j_2+1$ the values taken by $r$ are
$j_1\,-\,j_2,\,j_1\,-\,j_2\,+{1\over 2},\cdots,\,j_1+j_2$.
Remarkably these are the isospin values that are obtained when one
performs the tensor product of two ${\rm osp}(1\vert 2)$
irreducible representations of isospins $j_1$ and $j_2$ [\Pais]. It is
interesting to compare this result with the $SU(2)$ case in which
only isospins $j_1\,-\,j_2,\,j_1\,-\,j_2\,+\,1\,,\cdots,\,j_1+j_2$
are obtained. 
 
The coefficients $N_p$ in \vcuatro\ are given  by
$4j_2$-dimensional integrals which are, in general, quite hard to
evaluate. For this reason let us consider from now on the case 
$j_2\,=\,1/2$. If we simply put $j_1\,=\,j$, we have in this case
three two-dimensional integrals to compute: 
$$
\eqalign{
N_1(j)\,=&\,-\int_1^{\infty}\,du_1\int_1^{u_1}du_2\,
(u_1u_2)^{a+b}\,(u_1-1)^b\,(u_2-1)^b\,(u_1\,-\,u_2)^{2\rho-1}\,
\times\cr\cr
&\times\Bigl[\,{2j\over u_1}\,+\,{1\over u_1-1}\,+\,
(u_1\leftrightarrow u_2)\,\Bigr]\cr\cr
N_2(j)\,=&\,-2j\int_1^{\infty}\,du_1\,u_1^{a+b+2\rho-1}(u_1-1)^b\,
\int_0^{1}\,dv_1\,v_1^{a-1}\,(1-v_1)^b\cr\cr
N_3(j)\,=&\,-2j\int_0^{1}\,dv_1\int_0^{v_1}dv_2\,
v_1^{a}\,v_2^a\,(1-v_1)^b\,(1-v_2)^b\,(v_1-v_2)^{2\rho-1}
\,\Bigl[\,{1\over v_1}\,+\,(v_1\leftrightarrow v_2)\,\Bigr].\cr}
\eqn\vcinco
$$
The integral for $N_2(j)$ is easy to calculate since it factorizes
into two one dimensional integrals which are given in terms of the
Euler $\Gamma$-functions. The result is:
$$
N_2(j)\,=\,-2j\,{\Gamma(-a-2b-2\rho)\Gamma(1+b)\over 
\Gamma(-a-b-2\rho+1)}\,
{\Gamma(a)\Gamma(1+b)\over 
\Gamma(1+a+b)}\,.
\eqn\vseis
$$
Moreover, the quantities $N_1(j)$ and $N_3(j)$ 
can be put in terms of the
basic integrals of the Selberg type:
$$
J(\alpha,\beta,\gamma)\,\equiv\,
\int_0^{1}\,du_1\int_0^{u_1}du_2\,
[\,u_1^{\alpha+1}u_2^{\alpha}\,+\,(u_1\leftrightarrow u_2)\,]\,
(1-u_1)^{\beta}(1-u_2)^{\beta}
(u_1\,-\,u_2)^{2\gamma}\,.
\eqn\vsiete
$$
In fact, after some simple manipulations one can write:
$$
\eqalign{
N_1(j)\,=&\,J(-1-a-2b-2\rho, b, \rho-{1\over 2})\cr
N_3(j)\,=&\,-2jJ(a-1, b, \rho-{1\over 2})\,.\cr}
\eqn\vocho
$$
The function $J(\alpha,\beta,\gamma)$ is easily obtained from the
integrals evaluated in ref. [\DF]. One finds:
$$
J(\alpha,\beta,\gamma)\,=\,2\,
{\Gamma(2\gamma)\over\Gamma(\gamma)}\,
{\Gamma(1+\alpha)\Gamma(1+\beta)\over 
\Gamma(2+\alpha+\beta+\gamma)}\,
{\Gamma(2+\alpha+\gamma)\Gamma(1+\beta+\gamma)\over 
\Gamma(3+\alpha+\beta+2\gamma)}\,.
\eqn\vnueve
$$
Let us now see how one can get the operator algebra of the theory
from our results. First of all, it is clear from eqs. \vtres\ and
\vcuatro\ that the coefficients appearing in the operator algebra
must be related to the constants:
$$
S_p(j)\,=\,X_p\,(N_p(j))^2\,.
\eqn\treinta
$$
Indeed $S_p(j)$ is the coefficient of the non-analytic singularity
corresponding to the channel $p$ in the physical correlator 
$G(z\,,\,\bar z)$. Moreover the general form of the constants
$X_p$ has been obtained in ref. [\DF] from the integral
representation of the blocks. Adapting this result to our case, we
can write:
$$
\eqalign{
X_p\,=&\,\prod_{i=1}^{p-1}\,s(i(\rho-{1\over 2})\,)\,
\prod_{i=0}^{p-2}\,
{s(a+i(\rho-{1\over 2})\,)\,s(1+b+i(\rho-{1\over 2})\,)\over
s(1+a+b+(p-2+i)(\rho-{1\over 2})\,)}\,\times\cr\cr
\times&\,\prod_{i=1}^{3-p}\,s(i(\rho-{1\over 2})\,)\,
\prod_{i=0}^{2-p}\,
{s(1-a-2b-2\rho+i(\rho-{1\over 2})\,)\,
s(1+b+i(\rho-{1\over 2})\,)\over
s(1-a-b+(i-p)(\rho-{1\over 2})\,)},\cr}
\eqn\tuno
$$
where $s(x)\,\equiv\,{\rm sin}\,(\pi x)$. 

The operator algebra of the theory has the general form:
$$
\Phi_{m_1}^{j_1}(z_1,\bar z_1)\,\Phi_{m_2}^{j_2}(z_2,\bar z_2)\,=\,
\sum_{r,m}\,D_{j_1,m_1;j_2,m_2}^{r,m}\,\,\Bigl[
\,{\Phi_{m}^{r}(z_2,\bar z_2)\over 
|z_1\,-z_2|^{2(\Delta_{j_1}\,+\,\Delta_{j_2}\,-\,\Delta_{r})}}\,+\,
O(z_1\,-z_2)\,\Bigr].
\eqn\tdos
$$
In order to extract the structure constants 
$D_{j_1,m_1;j_2,m_2}^{r,m}$ of the theory from the quantities
$S_p(j)$ one must conveniently normalize the later. To fix this
normalization let us recall that the two-point functions of the
model must be normalized as:
$$
<\,\Phi_{m_1}^{j_1}(z_1,\bar z_1)\,\Phi_{m_2}^{j_2}(z_2,\bar z_2)\,>
\,={\delta_{j_1,j_2}\,\delta_{m_1,-m_2}\over
|z_1\,-z_2|^{4\Delta_{j_1}}}\,,
\eqn\ttres
$$
which implies the following constraint for the structure constants:
$$
D_{j_1,m_1;j_1,-m_1}^{0,0}\,=\,1\,.
\eqn\tcuatro
$$
Let us now see how one can implement this constraint in our
formalism. The condition \tcuatro\ is automatically fulfilled if we
divide the coefficients $S_p(j)$ by the one that corresponds to
the unit operator in the intermediate state. Clearly the
s-channel isospin $r$ is equal to zero only when $j=1/2$ and
$p=3$(recall that $j\geq 1/2$). Applying this reasoning to the 
$r=j\pm{1\over 2}$ cases, one is led to write:
$$
\Bigl[\,D_{j,j;{1\over 2},-{1\over 2}}
^{j-{1\over 2},j-{1\over 2}}\,\Bigr]^2\,=
\,{S_3(j)\over S_3({1\over 2})}\,,
\,\,\,\,\,\,\,\,\,\,\,\,\,\,\,\,\,\,\,\,\,\,\,
\Bigl[\,D_{j,j;{1\over 2},-{1\over 2}}
^{j+{1\over 2},j-{1\over 2}}\,\Bigr]^2\,=
\,{S_1(j)\over S_3({1\over 2})}\,.
\eqn\tcinco
$$
Using eqs. \vseis, \vocho, \vnueve\ and \tuno\ and the relation 
$s(x)\,\Gamma^2(x)\,=\,\pi\Gamma(x)/\Gamma(1-x)$, the structure
constants of eq. \tcinco\ are found to be:
$$
\eqalign{
&\Bigl[\,D_{j,m_1;{1\over 2},m_2}
^{j-{1\over 2},m_1+m_2}\,\Bigr]^2\,=\,
\Bigl[\,C_{j,m_1;{1\over 2},m_2}
^{j-{1\over 2},m_1+m_2}\,\,\Bigr]^4\,\,\,
{\Gamma({1\over 2}\,+\,\rho)\,\Gamma({1\over 2}\,-\,3\rho)\over
\Gamma({1\over 2}\,-\,\rho)\,\Gamma({1\over 2}\,+\,3\rho)}
{\Gamma({1\over 2}\,+\,(4j+1)\rho)\,
\Gamma({1\over 2}\,-\,(4j-1)\rho)\over
\Gamma({1\over 2}\,-\,(4j+1)\rho)\,
\Gamma({1\over 2}\,+\,(4j-1)\rho)}\cr\cr
&\Bigl[\,D_{j,m_1;{1\over 2},m_2}
^{j+{1\over 2},m_1+m_2}\,\Bigr]^2\,=\,
\Bigl[\,C_{j,m_1;{1\over 2},m_2}
^{j+{1\over 2},m_1+m_2}\,\,\Bigr]^4\,\,\,
{\Gamma({1\over 2}\,+\,\rho)\,\Gamma({1\over 2}\,-\,3\rho)\over
\Gamma({1\over 2}\,-\,\rho)\,\Gamma({1\over 2}\,+\,3\rho)}\,
{\Gamma({1\over 2}\,+\,(4j+3)\rho)\,
\Gamma({1\over 2}\,-\,(4j+1)\rho)\over
\Gamma({1\over 2}\,-\,(4j+3)\rho)\,
\Gamma({1\over 2}\,+\,(4j+1)\rho)},\cr\cr}
\eqn\tseis
$$
where $C_{j,m_1;{1\over 2},m_2}^{r,m_1+m_2}$ 
are the ${\rm osp}(1\vert 2)$ Clebsch-Gordan coefficients. These
coefficients have been obtained in ref. [\Pais]. Actually only the
values:
$$
\eqalign{
C_{j,j;{1\over 2},-{1\over 2}}^{j-{1\over 2},j-{1\over 2}}\,=&\,
-1\cr
C_{j,j;{1\over 2},-{1\over 2}}^{j+{1\over 2},j-{1\over 2}}\,=&\,
-C_{j,j;{1\over 2},-{1\over 2}}^{j,j-{1\over 2}}\,=\,
{1\over \sqrt{2j+1}}\,,\cr}
\eqn\tsiete
$$
are needed in our present calculation.

It is interesting to point out that the state 
$|j\,,j-{1\over 2}\,>$ appearing in the tensor product 
$|j\,,j\,>\otimes |{1\over 2}\,,-{1\over 2}\,>$ has negative norm
and has to be normalized to $-1$. In fact this 
$|j\,,j-{1\over 2}\,>$ state has bosonic statistics which means that
it belongs to an odd representation, despite of the fact that we are
only considering products of even representations. These
considerations naturally lead us to define
$$
\Bigl[\,D_{j,j;{1\over 2},-{1\over 2}}^{j,j-{1\over 2}}\,\Bigr]^2\,=
\,-{S_2(j)\over S_3({1\over 2})}\,.
\eqn\tocho
$$
Again it is a simple exercise to work out the right-hand side of
\tocho\ and arrive at the following expression for the structure
constants of this channel:
$$
\eqalign{
\Bigl[\,D_{j,m_1;{1\over 2},m_2}^{j,m_1+m_2}\,\Bigr]^2\,=&\,
\Bigl[\,C_{j,m_1;{1\over 2},m_2}^{j,m_1+m_2}\,\Bigr]^4
\,2^{4-8\rho}\,\rho^2\,(2j)^2\,(2j+1)^2\,\,\,
{\Gamma({1\over 2}\,+\,\rho)\,\Gamma({1\over 2}\,-\,3\rho)\over
\Gamma({1\over 2}\,-\,\rho)\,\Gamma({1\over 2}\,+\,3\rho)}
\times\cr\cr
&\times
\Bigl[\,{\Gamma(1\,-\,\rho)\over
\Gamma(\rho)}\,\Bigr]^2\,\,
\Bigl[\,{\Gamma((4j+2)\rho)\,
\Gamma(-4j\rho)\over
\Gamma(1\,-\,(4j+2)\rho)\,
\Gamma(1+\,4j\rho)\,}\Bigr]^2,\cr}
\eqn\tnueve
$$
where we have used the duplication formula for the
$\Gamma$-function, 
$\sqrt{\pi}\,\Gamma(2z)=\,2^{2z-1}
\Gamma(z)\,\Gamma(z+{1\over 2})$, 
in order to arrive at the final result. 
It is easy to verify that the right-hand side of eq. \tnueve\ is
non-negative, which confirms the correctness of our prescription
\tocho.

Notice that we have only computed the structure constants of eqs.
\tseis\ and \tnueve\ for a particular case of $m_1$ and $m_2$. The
operator algebra  constants for different values of these numbers
can be obtained by studying some other correlators. One can, for
example, analyze the vacuum expectation values  
$<\,\Phi^{j}_{-j}(0)\,\Phi^{{1\over 2}}_{-m}(z)\,
\Phi^{ {1\over 2}}_{m}(1)\,\tilde\Phi^{j}_{j}(\infty)\,
Q^{n}\,>$ for $m=0,1/2$ and \break
$<\Phi^{j}_{-j+{1\over 2}}(0)\,\Phi^{{1\over 2}}_{-m}(z)\,
\Phi^{ {1\over 2}}_{m}(1)\,\tilde\Phi^{j}_{j-{1\over 2}}(\infty)\,
Q^{n}\,>$ for $m=0,\pm 1/2$. In this last case the explicit form
of the the conjugate operator $\tilde\Phi_{j-1/2}^j$ (see eq.
\quince) will be needed. These calculations can be done by the same
method we have followed for the four point function \dseis. The
results for the structure constants coincide with eqs. \tseis\ and
\tnueve.

Summarizing, we have been able to study the general structure of
the four point functions of the ${\rm osp}(1\vert 2)$ current
algebra by means of a free field representation. The formalism
allows to obtain the operator algebra of the model, as we have
illustrated in the particular case in which one of the primary
fields that are multiplied carries the quantum numbers of the
fundamental representation. Much work remains to be done. First of
all, it would be interesting to obtain the operator algebra for
the case of two arbitrary isospins. In order to achieve this
purpose one must  tackle the combinatorics and the
multiple integrals that one has to deal with in this general case.
Another interesting problem is the study of the quantum
hamiltonian reduction in this model. We hope that our results
might help to shed light on the relation between the ${\rm
osp}(1\vert 2)$ current algebras and the minimal superconformal
models at the level of correlators.

\ack
We are grateful to J.M.F. Labastida and P.M. Llatas for
discussions. This work was supported in part by DGICYT under
grant PB93-0344, and by CICYT under grant  AEN96-1673.

\refout

\end